# Determination of intrinsic damping of perpendicularly magnetized ultrathin films from time resolved precessional magnetization measurements

Amir Capua[1,*], See-hun Yang[1], Timothy Phung[1], Stuart S. P. Parkin[1,2]

[1] IBM Research Division, Almaden Research Center, 650 Harry Rd., San Jose, California 95120, USA
[2] Max Planck Institute for Microstructure Physics, Halle (Saale), D-06120, Germany

*e-mail: acapua@us.ibm.com

**Abstract:**

Magnetization dynamics are strongly influenced by damping, namely the loss of spin angular momentum from the magnetic system to the lattice. An "effective" damping constant $\alpha_{eff}$ is often determined experimentally from the spectral linewidth of the free induction decay of the magnetization after the system is excited to its non-equilibrium state. Such an $\alpha_{eff}$, however, reflects both intrinsic damping as well as inhomogeneous broadening that arises, for example, from spatial variations of the anisotropy field. In this paper we compare measurements of the magnetization dynamics in ultrathin non-epitaxial films having perpendicular magnetic anisotropy using two different techniques, time-resolved magneto optical Kerr effect (TRMOKE) and hybrid optical-electrical ferromagnetic resonance (OFMR). By using an external magnetic field that is applied at very small angles to the film plane in the TRMOKE studies, we develop an explicit closed-form analytical expression for the TRMOKE spectral linewidth and show how this can be used to reliably extract the intrinsic Gilbert damping constant. The damping constant determined in this way is in excellent agreement with that determined from the OFMR method on the same samples. Our studies indicate that the asymptotic high-field approach that is often used in the TRMOKE method to distinguish the intrinsic damping from the effective damping may result in significant error, because such high external magnetic fields are required to make this approach valid that they are out of reach. The error becomes larger the lower is the intrinsic damping constant, and thus may account for the anomalously high damping constants that are often reported in TRMOKE studies. In conventional ferromagnetic resonance (FMR) studies, inhomogeneous contributions can be readily distinguished from intrinsic damping contributions from the magnetic field dependence of the FMR linewidth. Using the analogous approach, we show how reliable values of the intrinsic damping can be extracted from TRMOKE in two distinct magnetic systems with significant perpendicular magnetic anisotropy: ultrathin CoFeB layers and Co/Ni/Co trilayers.

I. Introduction

Spintronic nano-devices have been identified in recent years as one of the most promising emerging technologies for future low power microelectronic circuits[1, 2]. In the heart of the dynamical spin-state transition stands the energy loss parameter of the Gilbert damping. Its accurate determination is of paramount importance as it determines the performance of key building blocks required for spin manipulation such as the switching current threshold of the spin transfer torque magnetic tunnel junction (MTJ) used in magnetic random access memory (MRAM) as well as the skyrmion velocities and the domain wall motion current threshold. Up-scaling for high logic and data capacities while obtaining stability with high retention energies require in addition that large magnetic anisotropies be induced. These cannot be achieved simply by engineering the geometrical asymmetries in the nanometer-scale range, but rather require harnessing the induced spin-orbit interaction at the interface of the ferromagnetic film to obtain perpendicular magnetic anisotropy (PMA)[2]. Hence an increasing effort is invested in the quest for perpendicular magnetized materials having large anisotropies with low Gilbert damping[3-11].

Two distinct families of experimental methods are typically used for measurement of Gilbert damping, namely, time-resolved pump-probe and continuous microwave stimulated ferromagnetic resonance (FMR), either of which can be implemented using optical and/or electrical methods. While in some cases good agreement between these distinct techniques have been reported[12, 13], there is often significant disagreement between the methods[14, 15].

When the time resolved pump-probe method is implemented using the magneto optical Kerr effect (TRMOKE), a clear advantage over the FMR method is gained in the ability to operate at very high fields and frequencies[16, 17]. On the other hand, the FMR method allows operation over a wider range of geometrical configurations. The fundamental geometrical restriction of the TRMOKE comes from the fact that the magnetization precessions are initiated from the perturbation of the effective anisotropy field by the pump pulse, by momentarily increasing the lattice temperature[18, 19]. In cases where the torque exerted by the effective anisotropy field is negligible, the pump pulse cannot sufficiently perturb the magnetization. Such a case occurs for example whenever the magnetization lays in the plane of the sample in uniaxial thin films having perpendicular magnetic anisotropy. Similar limitations exist if the magnetic field is applied perpendicular to the film. Hence in TRMOKE experiments, the external field is usually applied at angles typically not smaller than about $10°$ from either the film plane or its normal. This fact has however the consequence that the steady state magnetization orientation, determined by the balancing condition for the torques, cannot be described using an explicit-form algebraic expression, but rather a numerical approach should be taken[5]. Alternatively, the dynamics can be described using an effective damping from which the intrinsic damping, or at least an upper bound of its value, is estimated at the high magnetic field limit with the limit being undetermined. These approaches are hence less intuitive while the latter does not indicate directly on the energy losses but rather on the combination of the energy loss rate, coherence time of the spin ensemble and geometry of the measurement.

In this paper, we present an approach where the TRMOKE system is operated while applying the magnetic field at very small angles with respect to the sample plane. This enables us to use explicit closed-form analytical expressions derived for a perfectly in-plane external magnetic field as an approximate solution. Hence, extraction of the intrinsic Gilbert damping using an analytical model becomes possible without the need to drive the system to the high magnetic field limit providing at the same time an intuitive understanding of the measured responses. The validity of the method is verified using a highly sensitive hybrid optical-electrical FMR system (OFMR) capable of operating with a perfectly in-plane magnetic field where the analytical expressions hold. In particular, we bring to test the high-field asymptotic approach used for evaluation of the intrinsic damping from the effective damping and show that in order for it to truly indicate the intrinsic damping, extremely high fields need to be applied. Our analysis reveals the resonance frequency dispersion relation as well as the inhomogeneous broadening to be the source of this requirement which becomes more difficult to fulfill the smaller the intrinsic damping is. The presented method is applied on two distinct families of technologically relevant perpendicularly magnetized systems; CoFeB[4, 6] and Co/Ni/Co[20-23]. Interestingly, the results indicate that the Ta seed layer thickness used in CoFeB films strongly affects the intrinsic damping, while the static characteristics of the films remain intact. In the Co/Ni/Co trilayer system which has in contrast a large effective anisotropy field, unexpectedly large spectral linewidths are measured when the external magnetic field is comparable to the effective anisotropy field, which cannot be explained by the conventional model of non-interacting spins describing the inhomogeneous broadening. This suggests that under the low stiffness conditions associated with such bias fields, cooperative exchange interactions, as two magnon scattering, become relevant[8, 24].

## II. EXPERIMENT

The experiments presented were carried out on three PMA samples: two samples consisted of $Co_{36}Fe_{44}B_{20}$ which differed by the thickness of the underlayer and a third sample consisting of Co/Ni/Co trilayer. The CoFeB samples were characterized by low effective anisotropy ($H_{keff}$) values as well as by small distribution of its value in contrast to the Co/Ni/Co trilayer system. We define here $H_{keff}$ as $2K_u/M_s-4\pi M_s$ where $K_u$ is the anisotropy energy constant and $M_s$ being the saturation magnetization.

The structures of the two CoFeB samples were 50Ta|11CoFeB|11MgO|30Ta, and 100Ta|11CoFeB|11MgO|30Ta (units are in Å) and had similar $M_s$ values of 1200 emu/cc and $H_{keff}$ of 1400 Oe and 1350 Oe respectively. The third system studied was 100AlO$_x$|20TaN|15Pt|8Pt$_{75}$Bi$_{25}$|3Co|7Ni|1.5Co|50TaN with $M_s$ of 600 emu/cc and $H_{keff}$ value of about 4200 Oe. All samples were grown on oxidized Si substrates using DC magnetron sputtering and exhibited sharp perpendicular switching characteristics. The samples consisting of CoFeB were annealed for 30 min at 275° C in contrast to the Co/Ni/Co which was measured as deposited. Since the resultant film has a polycrystalline texture, the in-plane anisotropy is averaged out and the films are regarded as uniaxial crystals with the symmetry axis being perpendicular to the film plane.

The two configurations of the experimental setup were driven by a Ti:Sapphire laser emitting 70 fs pulses at 800 nm having energy of 6 nJ. In the first configuration a standard polar pump-probe TRMOKE was implemented with the probe pulse being attenuated by 15 dB compared to the pump pulse. Both beams were focused on the sample to an estimated spot size of 10.5 μm defined by the full width at half maximum (FWHM). In the hybrid optical-electrical OFMR system, the Ti:Sapphire laser served to probe the magnetization state via the magneto-optical Kerr effect after being attenuated to pulse energies of about 200 pJ and was phase-locked with a microwave oscillator in a similar configuration to the one reported in Ref. [25]. For this measurement, the film was patterned into a 20 μm x 20 μm square island with a Au wire deposited in proximity to it, which was driven by the microwave signal. Prior to reaching the sample, the probing laser beam traversed the optical delay line that enabled mapping of the time axis and in particular the out of plane-$m_z$ component of the magnetization as in the polar TRMOKE experiment. With this configuration the OFMR realizes a conventional FMR system where the magnetization state is read in the time-domain using the magneto optical Kerr effect and hence its high sensitivity. The OFMR system therefore enables operation even when the external field is applied fully in the sample plane.

## III. RESULTS AND DISCUSSION

### A. TRMOKE measurements on 50 Å-Ta CoFeB film

The first experiments we present were performed on the 50 Å-Ta CoFeB system which is similar to the one studied in Ref. [4]. The TRMOKE measurement was carried out at two angles of applied magnetic field, $\psi_H$, of $4°$ and $1°$ measured from the surface plane as indicated in Fig. 1. We define here in addition the complementary angle measured from the surface normal, $\theta_H = \pi/2 - \psi_H$. Having its origin in the effective anisotropy, the torque generated by the optical pump is proportional to $M_s H_{keff} \cos(\theta)\sin(\theta)$ with $\theta$ being the angle of the magnetization relative to the normal of the sample plane. Hence, for $\psi_H < 1°$, the angle $\theta$ becomes close to $\pi/2$, and the resultant torque generated by the optical pump is not strong enough to initiate reasonable precessions. For the same reason, the maximum field measureable for the $\psi_H = 1°$ case is significantly lower than for the $\psi_H = 4°$ case. This is clearly demonstrated in the measured MOKE signals for the two $\psi_H$ angles in Fig. 2(a). While for $\psi_H = 4°$ the precessional motion is clearly seen even at a bias field of 12 kOe, with $\psi_H = 1°$ the precessions are hardly observable already at a bias field of 5.5 kOe. Additionally, it is also possible that the lower signal to noise ratio observed for $\psi_H = 1°$ may be due to a breakdown into domains with the almost in-plane applied magnetic field[26]. After reduction of the background signal, the measured data can be fitted to a decaying sinusoidal response from which the frequency and decay time can be extracted in the usual manner [6] (Fig. 2(b)). The measured precession frequency as a function of the applied external field, $H_0$, is plotted in Fig. 3(a). Significant differences near $H_{keff}$ are observed for merely a change of three degrees in the angle of the applied magnetic field. In particular, the trace for $\psi_H = 1°$ exhibits a minimum point at approximately $H_{keff}$ in contrast to the monotonic behavior of the $\psi_H = 4°$ case. The theoretical dependence of the resonance frequency on the magnetic bias field expressed in normalized units, $\omega/\gamma H_{keff}$, with $\omega$ being the resonance angular frequency and $\gamma$ the gyromagnetic ratio, is presented in Fig. 3(b) for several representative angles of the applied field. The resonance frequency at the vicinity of $H_{keff}$ is very sensitive to slight changes in the angle of the applied field as observed also in the experiment. Actually the derivative of the resonance frequency with respect to the applied field at the vicinity of $H_{keff}$ is even more sensitive where it diverges for $\theta = 90°$ but reaches a value of zero for the slightest angle divergence. A discrepancy between the measurement and the theoretical solution exists however. At field values much higher than $H_{keff}$ the precession frequency should be identical for all angles (Fig. 3(b)) but in practice the resonance frequency measured for $\psi_H$ of $4°$ is consistently higher by nearly 2 GHz than at $1°$. The theory also predicts that for the case of $4°$, the resonance frequencies should exhibit a minimum point as well which is not observed in the measurement. The origin of the difference is not clear and may be related to the inhomogeneities in the local fields or to the higher orders of the interface induced anisotropy which were neglected in the theoretical calculation.

In Fig. 3(c), we plot the effective Lorentzian resonance linewidth in the frequency domain, $\Delta\omega_{eff}$, defined by $\Delta\omega_{eff} = 2/\tau_{eff}$ with $\tau_{eff}$ being the measured decay time extracted from the measured responses. Decomposing the measured linewidth to an intrinsic contribution that represents the energy losses upon precession and an extrinsic contribution which represents the inhomogeneities in the local fields and is not related to energy loss of the spin system, we express the linewidth as: $\Delta\omega_{eff} = \Delta\omega_{int} + \Delta\omega_{IH}$. $\Delta\omega_{int}$ is given by the Smit-Suhl formula[27, 28] and equals $2/\tau$ with $\tau$ denoting the intrinsic spin precession decay time whereas $\Delta\omega_{IH}$ represents the dispersion in the resonance frequencies due to the inhomogeneities. If the variations in the resonance frequency are assumed to be primarily caused by variations in the local effective anisotropy field $\Delta H_{keff}$, $\Delta\omega_{IH}$ may be given by: $\Delta\omega_{IH} = |d\omega/dH_{keff}| \cdot \Delta H_{keff}$. For the case of $\theta_H = \pi/2$ or $\theta_H = 0$, $\Delta\omega_{eff}$ has a closed mathematical form. In PMA films with bias field applied in the sample plane, the expression for $\Delta\omega_{eff}$ becomes:

$$\Delta\omega_{eff} = \alpha\gamma\left(2H_0 - H_{keff}\right) + \frac{\gamma H_0}{2\sqrt{H_0^2 - H_0 H_{keff}}} \Delta H_{keff} \quad \text{for} \quad H_0 > H_{keff}$$

$$\Delta\omega_{eff} = \alpha\gamma H_0\left(\frac{2H_{keff}}{H_0} - \frac{H_0}{H_{keff}}\right) + \frac{\gamma H_{keff}}{\sqrt{H_{keff}^2 - H_0^2}} \Delta H_{keff} \quad \text{for} \quad H_0 < H_{keff}$$

, (1)

with $\alpha$ denoting the Gilbert damping. The first terms in Eq. (1) stem from the intrinsic damping, while the second terms stem from the inhomogeneous broadening. Eq. (1) shows that while the contribution of the intrinsic part to the total spectral linewidth is finite, as the external field approaches $H_{keff}$ either from higher or lower field values, the inhomogeneous contribution diverges. Equation (1) further shows that for $H_0 >> H_{keff}$, the slope of $\Delta\omega_{eff}$ becomes $2\alpha\gamma$ with a constant offset given by $\gamma\Delta H_{keff}/2$. Although Eq. (1) is valid only for $\theta_H = \pi/2$, it is still instructive to apply it on the measured linewidth for the $\psi_H = 4°$ case.

The theoretical intrinsic linewidth for $\theta_H = \pi/2$, inhomogeneous contribution and the sum of the two after fitting $\alpha$ and $\Delta H_{keff}$ in the range $H_0 > 5000$ Oe are plotted in Fig. 3(c). The resultant fitting values were 0.023±0.002 for the Gilbert damping and 175 Oe for $\Delta H_{keff}$. At external fields comparable to $H_{keff}$ the theoretical expression derived for the inhomogeneous broadening for a perfectly in-plane field does not describe properly the experiment. In the theoretical analysis, at fields comparable to $H_{keff}$, the derivative $d\omega/dH_0$ diverges and therefore also the derivative $d\omega/dH_{keff}$ as understood from Fig. 3(b). In the experiment however, $\theta_H \neq \pi/2$ and the actual derivative $d\omega/dH_{keff}$ approaches zero. Hence any variation in $H_{keff}$ results in minor variation of the frequency. This means that the contribution of the inhomogeneous broadening to the total linewidth is suppressed near $H_{keff}$ in the experiment as opposed to being expanded in the theoretical calculation which was carried out for $\theta_H = \pi/2$. The result is an overestimated theoretical linewidth near $H_{keff}$. After reduction of the inhomogeneous broadening, the extracted intrinsic measured linewidth is presented in Fig. 3(c) as well showing the deviation from the theoretical intrinsic contribution as the field approaches $H_{keff}$.

To further investigate the effect of tilting the magnetic field, we study the TRMOKE responses for the $\psi_H = 1°$ case. The measured linewidth for this case is presented in Fig. 3(d). In contrast to the $\psi_H = 4°$ case, the measured linewidth now increases at fields near $H_{keff}$ as expected theoretically. Furthermore, the measured linewidth for the $\psi_H = 1°$ case is well described by Eq. (1) even in the vicinity of $H_{keff}$ as well as for bias fields smaller than $H_{keff}$. The fitting results in the same damping value of 0.023±0.0015 as with the $\psi_H = 4°$ case, and a variation in $\Delta H_{keff}$ of 155 Oe, which is 20 Oe smaller than the value fitted for the $\psi_H = 4°$ case.

We next turn to examine the Gilbert damping. In the absence of the demagnetization and crystalline anisotropy fields, the expression for the intrinsic Gilbert damping is given by:

$$\alpha = \frac{1}{\tau\omega}. \qquad (2)$$

Once the anisotropy and the demagnetization fields are included, the expression for the intrinsic Gilbert damping becomes:

$$\begin{array}{ll} \alpha = \gamma \left|\dfrac{dH_0}{d\omega}\right| \cdot \dfrac{1}{\tau\omega} & \text{for} \quad H_0 > H_{keff} \\[2ex] \alpha = \dfrac{2}{\left(2H_{keff}/H_0 - H_0/H_{keff}\right)} \cdot \gamma \left|\dfrac{dH_0}{d\omega}\right| \cdot \dfrac{1}{\tau\omega} & \text{for} \quad H_0 < H_{keff} \end{array}, \qquad (3)$$

and is valid only for $\theta_H = \pi/2$ and for crystals having uniaxial symmetry. At other angles a numerical method[5] should be used to relate the precession decay time to the Gilbert damping. Eq. (3) is merely the intrinsic contribution in Eq. (1) written in the form resembling Eq. (2). At high fields both Eqs. (2) and (3) converge to the same result since $\frac{dH_0}{d\omega} \to \gamma^{-1}$. As seen in Fig. 3(b), at bias fields comparable to $H_{keff}$ the additional derivative term of Eq. (3) becomes very significant. When substituting the measured decay time, $\tau_{eff}$, for $\tau$, Eq. (2) gives what is often interpreted as the "effective" damping, $\alpha_{eff}$, from which the intrinsic damping is measured by evaluating it at high fields when the damping becomes asymptotically field independent. Additionally, the asymptotic limit should be reached with respect to the inhomogeneous contribution of Eq. (1). In Fig. 3(e), we plot the effective damping using $\tau_{eff}$ and Eq. (2). We further show the intrinsic damping value after extracting the intrinsic linewidth and using Eq. (3). Examining first the effective damping values, we see that for the two angles, the values are distinctively different at low fields but converge at approximately 4100 Oe (Beyond 5500 Oe the data for the $\psi_H = 1°$ case could not be measured). In fact, the behavior of the effective damping seems to be related to the dependence of the resonance frequency on $H_0$ (Fig. 3(a)) in which for the $\psi_H = 1°$ case reaches an extremum while the $\psi_H = 4°$ case exhibits a monotonic behavior. Since Eq. (2) lacks the derivative term $|dH_0/d\omega|$, near $H_{keff}$ the effective damping is related to the Gilbert damping by the relation: $\alpha_{eff} = \frac{1}{\gamma}\frac{d\omega}{dH_0}\alpha$ for $H_0 > H_{keff}$. Furthermore, since $\alpha$ does not depend on the magnetic field to the first order, the dependence of the effective damping, $\alpha_{eff}$, on the bias field stems from the derivative term $|d\omega/dH_0|$ which becomes larger and eventually diverges to infinity when the magnetic field reaches $H_{keff}$ as can be inferred from Fig. 3(b) for the case of $\psi_H = 0°$ for which Eq. (3) was derived. Hence the increase in $\alpha_{eff}$ at bias fields near $H_{keff}$. The same considerations apply also for $H_0 < H_{keff}$. As the angle $\psi_H$ increases, this analysis becomes valid only for bias fields which are large enough or small enough relative to $H_{keff}$. When examined separately, each effective damping trace may give the impression that at the higher fields it has become bias field independent and reached its asymptotic value from which two very distinct Gilbert damping values of ~0.027 and ~0.039 are extracted at field values of 12 kOe and 5.5 kOe for the $\psi_H = 4°$ and $\psi_H = 1°$ measurements, respectively. These values are also rather different from the intrinsic damping value of 0.023 extracted using the analytical model. In contrast to the effective damping, the intrinsic damping obtained from the analytical model reveals a constant and continuous behavior which is field and angle independent. The presumably negative values measured for the $\psi_H = 4°$ case stem of course from the fact that the expressions in Eqs. (1) and (3) are derived for the $\theta_H = \pi/2$ case. The error in using the effective damping in conjunction with the asymptotic approximation compared to using the analytical model is therefore 17% and 70% for the $\psi_H = 4°$ and $\psi_H = 1°$ measurements respectively.

It is important in addition to understand the consequence of using Eq. (2) rather than Eq. (3). In Fig. 3(f) we present the error in the damping value after accounting for the inhomogeneous broadening using Eq. (2) instead of the complete expression of Eq. (3). As expected, the error increases as the applied field approaches $H_{keff}$. For the measurement taken with $\psi_H = 4°$ the error is significantly smaller due to the smaller value of the derivative $d\omega/dH_0$.

As mentioned previously, in order to evaluate the intrinsic damping from the total measured linewidth, the asymptotic limit should be reached with respect to the inhomogeneous broadening as well (Eq. (1)). In Figs. 3(c) and 3(d) we see that this is not the case where the contribution of the inhomogeneous linewidth is still large compared to the intrinsic linewidth. Examining Figs. 3(d) and 3(f) for the case of $\psi_H = 1°$, we see that the overall error of 70% resulting in the asymptotic evaluation stems from both the contribution of inhomogeneous broadening as well as from the use of Eq. (2) rather than Eq. (3) while for $\psi_H = 4°$ (Figs. 3(c) and 3(f)) the error of 17% is solely due to contribution of the inhomogeneous broadening which was not as negligible as conceived when applying the asymptotic approximation.

### B. Comparison of TRMOKE and OFMR measurements in 100 Å-Ta CoFeB film

We next turn to study the magnetization dynamics using the OFMR system where the precessions are driven with the microwave signal. Hence, the external magnetic field can be applied perfectly in the sample plane. The 100 Å-Ta CoFeB sample was used for this experiment. Before patterning the film for the OFMR measurement, a TRMOKE measurement was carried out at $\psi_H = 4°$ which exhibited a similar behavior to that observed with the sample having 50 Å Ta as a seeding layer. The dependence of the resonance frequency on the magnetic field as well as the measured linewidth and its different contributions are presented in Figs. 4(a) and 4(b). Before reduction of the inhomogeneous broadening the asymptotic effective damping was measured to be ~0.0168 while after extraction of the intrinsic damping a value of 0.0109±0.0015 was measured marking a difference of 54% (Fig. 4(c)). The fitted $\Delta H_{keff}$ was 205 Oe. Fig. 4(b) shows that the origin of the error stems from significant contribution of the inhomogeneous broadening compared to the intrinsic contribution which plays a more significant role when the damping is low. By using the criteria for the minimum field that results in $\Delta \omega_{IH} = \Delta \omega_{eff}/10$ to estimate the point where the asymptotic approximation would be valid, we arrive to a value of at least 4.6 T which is rather impractical. The threshold of this minimal field is highly dependent on the damping so that for a lower damping an even higher field would be required.

An example of a measured trace using the OFMR system at a low microwave frequency of 2.5 GHz is presented in Fig. 4(d). The square root of the magnetization amplitude (out of plane $m_z$ component) while preserving its sign is plotted to show detail. The high sensitivity of the OFMR system enables operation at very low frequencies and bias fields. For every frequency and DC magnetic field value, several cycles of the magnetization precession were recorded by scanning the optical delay line. The magnetic field was then swept to fully capture the resonance. The trace should be examined separately in two sections, below $H_{keff}$ and above $H_{keff}$ (marked in the figure by black dashed line). For frequencies of up to $\gamma H_{keff}$ two resonances are crossed as indicated by the guiding red dashed line which represents the out-of-phase component of the magnetization, namely the imaginary part of the magnetic susceptibility. Hence the cross section along this line gives the field dependent absorption spectrum from which the resonance frequency and linewidth can be identified. This spectrum is shown in Fig. 4(e) together with the fitted lorentzian lineshapes for bias fields below and above $H_{keff}$. The resultant resonance frequencies of all measurements are plotted in addition in Fig. 4(a).

The resonance linewidths extracted for bias fields larger than $H_{keff}$, are presented in Fig. 4(f). Here the effective magnetic field linewidth, $\Delta H_{eff}$, that includes the contribution of the inhomogeneous broadening derived from the same principles that led to Eq. (1) with $\theta_H = \pi/2$ is given by:

$$\Delta H_{eff} = \frac{2\alpha\omega}{\gamma} + \frac{1}{2}\left(1 + \frac{H_{keff}}{\sqrt{H_{keff}^2 + 4\left(\frac{\omega}{\gamma}\right)^2}}\right)\Delta H_{keff} \qquad \text{for} \qquad H_0 > H_{keff}$$

(5).

$$\begin{cases} \Delta H_{eff} = \frac{\alpha\omega}{\gamma}\left(\frac{2H_{keff}}{H_0} - \frac{H_0}{H_{keff}}\right) + \frac{H_{keff}}{H_0}\Delta H_{keff} & \text{for} \qquad H_0 < H_{keff} \\ \text{with} \qquad H_0 = \sqrt{H_{keff}^2 - (\omega/\gamma)^2} \end{cases}$$

The second terms in Eq. (5) denote the contribution of the inhomogeneous broadening, $\Delta H_{IH}$, and are frequency dependent as opposed to the case where the field is applied out of the sample plane[9]. The dispersion in the effective anisotropy, $\Delta H_{keff}$, and the intrinsic Gilbert damping were found by fitting the linewidth in the seemingly linear range at frequencies larger than 7.5 GHz. The contributions of the intrinsic and inhomogeneous parts and their sum are presented as well in Fig. 4(f).

It is apparent that the measured linewidth at the lower frequencies is much broader than the theoretical one. The reason for that lies in the fact that in practice the bias field is not applied perfectly in the sample plane as well as in the fact that there might be locally different orientations of the polycrystalline grains due to the natural imperfections of the interfaces that further result in angle distribution of $\theta_H$. Since the measured field linewidth is a projection of the spectral linewidth into the magnetic field domain, the relation between the frequency and the field intrinsic linewidths is given by: $\Delta H_{int} = \Delta \omega_{int} \cdot \left( \frac{d\omega}{dH_0} \right)^{-1}$. The intrinsic linewidth, $\Delta \omega_{int}$, in the frequency domain near $H_{keff}$ is finite, as easily seen from Eq. (1) while the derivative term near $H_{keff}$ is zero for even the slightest angle misalignment as already seen. Hence the field-domain linewidth diverges to infinity as observed experimentally. The inhomogeneous broadening component does not diverge in that manner but is rather suppressed. To show that the excessive linewidth at low fields is indeed related to the derivative of $d\omega/dH_0$ we empirically multiply the total theoretical linewidth by the factor $d\omega/d(\gamma H_0)$ which turns out to fit the data surprisingly well (Fig. 4(f)). This is merely a phenomenological qualitative description, and a rigorous description should still be derived.

The fitted linewidth of Fig. 4(f) results in the intrinsic damping value of 0.011±0.0005 and is identical to the value obtained by the TRMOKE method. Often concerns regarding the differences between the TRMOKE and FMR measurements such as spin wave emission away from the pump laser spot in the TRMOKE[29], increase of damping due to thermal heating by the pump pulse as well as differences in the nature of the inhomogeneous broadening are raised. Such effects do not seem to be significant here. Additionally, it is worth noting that since the linewidth seems to reach a linear dependence with respect to the field at high fields, it may be naively fitted using a constant frequency-independent inhomogeneous broadening factor. In that case an underestimated value of ~0.0096 would have been obtained. The origin of this misinterpretation is seen clearly by examining the inhomogeneous broadening contribution in Fig. 4(f) that shows it as well to exhibit a seemingly linear dependence at the high fields. Regarding the inhomogeneous broadening, the anisotropy field dispersion, $\Delta H_{K_{eff}}$, obtained with the TRMOKE was 205 Oe while the value obtained from the OFMR system was 169 Oe. Although these values are of the same order of magnitude, the difference is rather significant. It is possible that the discrepancy is related to the differences in the measurement techniques. For instance, the fact that both the pump and probe beams have the same spot size may cause an uneven excitation across the probed region in the case of the TRMOKE measurement while in the case of the OFMR measurement the amplitude of the microwave field decays at increasing distances away from the microwire. These effects may be reflected in the measurements as inhomogeneous broadening. Nevertheless, the measured intrinsic damping values are similar.

Finally, we compare the effective damping of the OFMR and the TRMOKE measurements without correcting for the inhomogeneous broadening in Fig. 4(g). The figure shows a deviation in the low field values which is by now understood to be unrelated to the energy losses of the system.

Furthermore, we observe that the thickness of the Ta underlayer affects the damping. The comparison of the 50 Å-Ta CoFeB and the 100 Å-Ta CoFeB samples shows that the increase by merely 50 Å of Ta, reduced significantly the damping while leaving the anisotropy field unaffected.

### C. TRMOKE and OFMR measurements in Co/Ni/Co film

In the last set of measurements we study the Co/Ni/Co film which has distinctively different static properties compared to the CoFeB samples. The sample was studied using the TRMOKE setup at two $\psi_H$ angles of $1°$ and $4°$ and using the OFMR system at $\psi_H = 0°$. The resultant resonance frequency traces are depicted in Fig. 5(a). The spectral linewidth measured for $\psi_H = 4°$ using the TRMOKE setup is presented in Fig. 5(b). A linear fit at the quasi linear high field range results in a large damping value of 0.081±0.015 and in a very large $\Delta H_{K_{eff}}$ of 630 Oe. The large damping is attributed to the efficient spin pumping into PtBi[30] layer having large spin-orbit coupling. When the angle of the applied magnetic field is reduced to $\psi_H = 1°$ a clearer picture of the contribution of the inhomogeneous broadening to the total linewidth is obtained (Fig. 5(c)) revealing that it cannot explain solely the measured spectral linewidths. While the theoretical model predicts that the increase in bandwidth spans a relatively narrow field range around $H_{keff}$, the measurement shows an increase over a much larger range around $H_{keff}$. The linewidth broadening originating from the anisotropy dispersion was theoretically calculated under the assumption of a small perturbation of the resonance frequency. A large $\Delta H_{keff}$ value was measured however from the TRMOKE measurement taken at $\psi_H = 4°$. Calculating numerically the exact variation of the resonance frequency improved slightly the fit but definitely did not resolve the discrepancy (not presented). From this fact we understand that there should be an additional source contributing to the line broadening at least near $H_{keff}$. A possible explanation may be related to the low stiffness[27] associated with the $H_0 \approx H_{keff}$ conditions. Under such conditions weaker torques which are usually neglected may become relevant[24, 31]. These torques could possibly originate from dipolar or exchange coupling resulting in two magnon scattering processes or even in a breakdown into magnetic domains as described by Grolier et al.[26]. From the limited data range at this angle, the damping could not be measured.

The OFMR system enabled a wider range of fields and frequencies than the ones measured with the TRMOKE for $\psi_H = 1°$ (Fig. 5(a)). Fig. 5(d) presents the measured OFMR linewidth. The quasi-linear regime of the linewidth seems to be reached at frequencies of 12 GHz corresponding to bias field values which are larger than 7500 Oe. The resultant intrinsic damping after fitting to this range was 0.09±0.005 with a $\Delta H_{keff}$ of 660 Oe which differ by approximately 10% from the values obtained from the TRMOKE measurement. The effective measured damping is plotted in Fig. 5(e). The asymptotic damping value, though not fully reached for this high damping sample, would be about 0.1.

This represents an error of about 10% which is smaller compared to the errors of 17% and 54% encountered in the CoFeB samples because of the larger damping of the Co/Ni/Co sample.

### D. Considerations of two-magnon scattering

In general, two-magnon spin wave scattering by impurities may exist in our measurements at all field ranges[32, 33], not only near $H_{keff}$ as suggested in the discussion of the previous section[32, 33]. The resultant additional linewidth broadening would then be regarded as an extrinsic contribution to the damping[34-36]. While in isotropic films which exhibit low crystalline anisotropy or in films having in-plane crystalline anisotropy, two-magnon scattering is maximized when the external field is applied in the film plane, in PMA films this is not necessarily the case and the highest efficiency of two-magnon scattering may be obtained at some oblique angle[35].

In films where two-magnon scattering is significant, the measured linewidth should exhibit an additional nonlinear dependence on the external field which cannot be accounted for by the present model. In such case, a strong dependence on the external field would be observed for fields below $H_{keff}$ due to the variation in the orientation of the magnetization with the external magnetic field. At higher fields the dependence on the external field is expected to be moderate[35].

While at bias field values below $H_{keff}$ our data is relatively limited, at external magnetic fields that are larger than $H_{keff}$, the observed linewidth seems to be described well by our model resulting in a field independent Gilbert damping coefficient. This seems to support our model that the scattering of spin waves does not have a prominent effect. It is possible however that a moderate dependence on the bias field, especially at high field values, may have been "linearized" and classified as intrinsic damping.

## IV.   CONCLUSION

In conclusion, in this paper we studied the time domain magnetization dynamics in non-epitaxial thin films having perpendicular magnetic anisotropy using the TRMOKE and OFMR systems. The analytical model used to interpret the magnetization dynamics from the TRMOKE responses indicated that the asymptotic high-field approach often used to distinguish the intrinsic damping from the effective damping may result in significant error that increases the lower the damping is. Two sources for the error were identified while validity of the asymptotic approach was shown to require very high magnetic fields. Additionally, the effective damping was shown to be highly affected by the derivative of the resonance frequency with respect to the magnetic field $|d\omega/dH_0|$. The analytical approach developed here was verified by use of the OFMR measurement showing excellent agreement whenever the intrinsic damping was compared and ruled out the possibility of thermal heating by the laser or emission of spin waves away from the probed area.

As to the systems studied, a large impact of the seed layer on the intrinsic damping with minor effect on the static characteristics of the CoFeB system was observed and may greatly aid in engineering the proper materials for the MTJ. Interestingly, the use of the analytical model enabled identification of an additional exchange torque when low stiffness conditions prevailed. While effort still remains to understand the limits on the angle of the applied magnetic field to which the analytical solution is valid, the approach presented is believed to accelerate the discovery of novel materials for new applications.

# Acknowledgments:

A.C. thanks the Viterbi foundation and the Feder Family foundation for supporting this research.

Figure 1

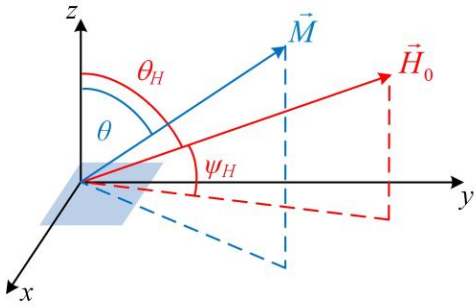

FIG 1. Illustration of the angles $\psi_H$, $\theta_H$ and $\theta$. M and H$_0$ vectors denote the magnetization and external magnetic field, respectively.

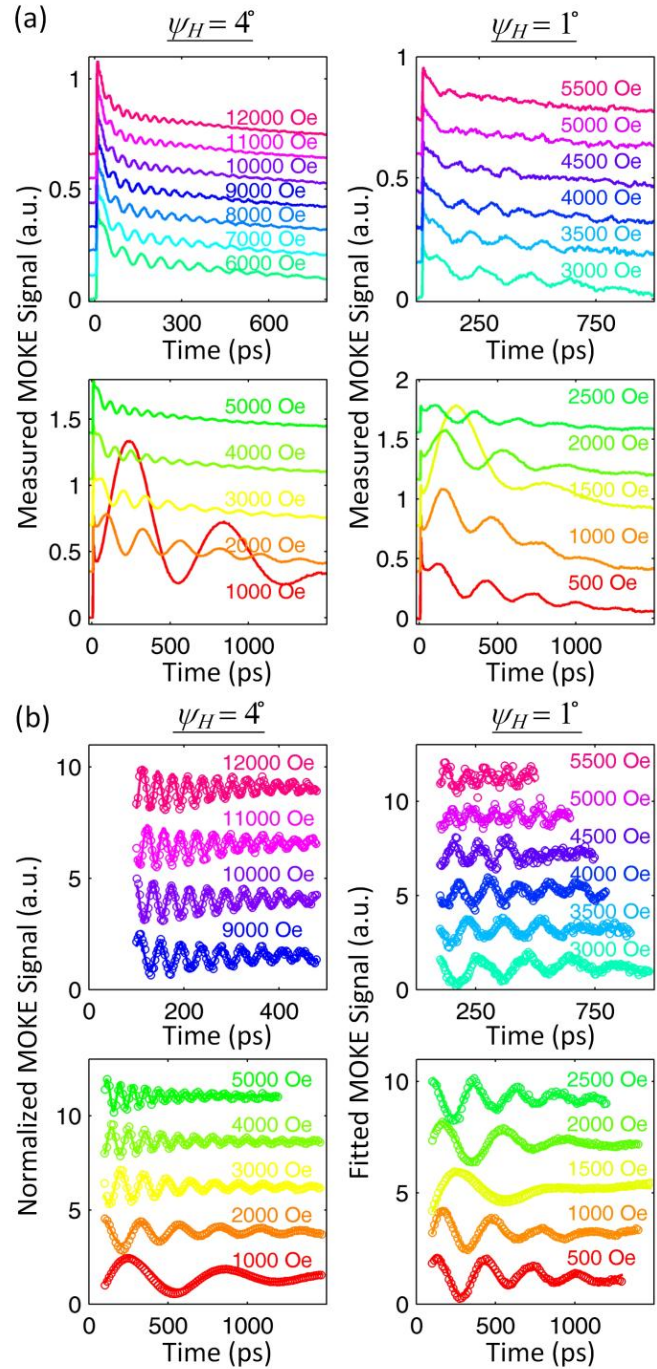

FIG. 2. Measured TRMOKE responses at $\psi_H$ angles of $4°$ and $1°$. (a) TRMOKE signal at low and high external magnetic field values. Traces are shifted for clarity. (b) Measured magnetization responses after reduction of background signal (open circles)

superimposed with the fitted decaying sine wave (solid lines). Traces are shifted and normalized to have the same peak amplitude. Data presented for low and high external magnetic field values.

Figure 3

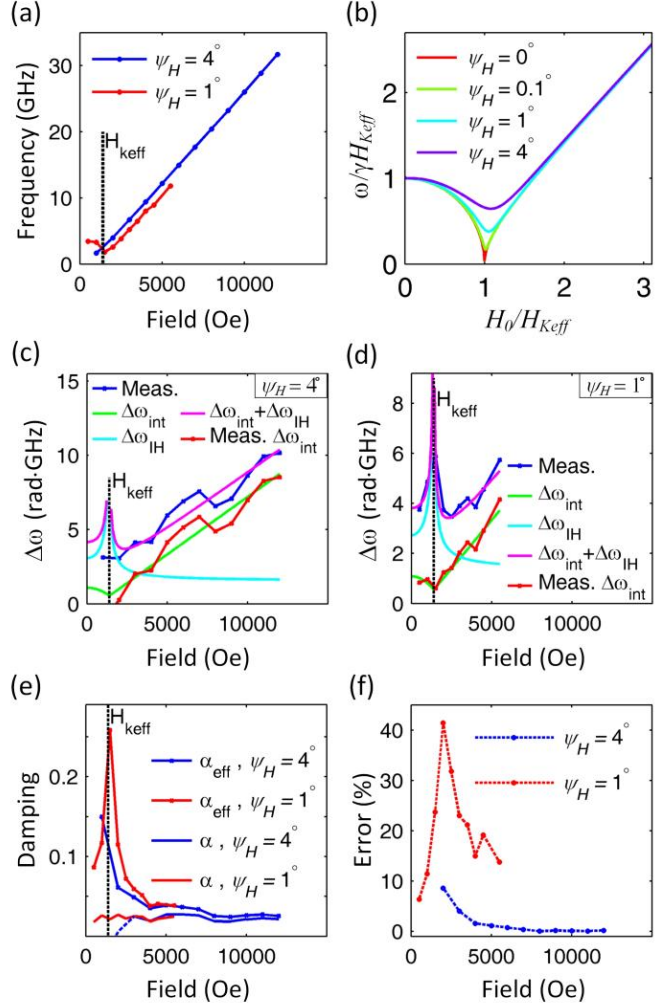

FIG. 3. TRMOKE measurements at $\psi_H = 4°$ and $\psi_H = 1°$. (a) Measured resonance frequency versus magnetic field. (b) Theoretical dependence of resonance frequency on magnetic field presented in normalized units. (c) & (d) Measured linewidth (blue), fitted theoretical contributions to linewidth (green, cyan, magenta) and extracted intrinsic linewidth from measurement (red) for $\psi_H = 4°$ and $\psi_H = 1°$, respectively. (e) Intrinsic and effective damping. (f) Error in damping value when using Eq. (2) instead of Eq. (3).

Figure 4

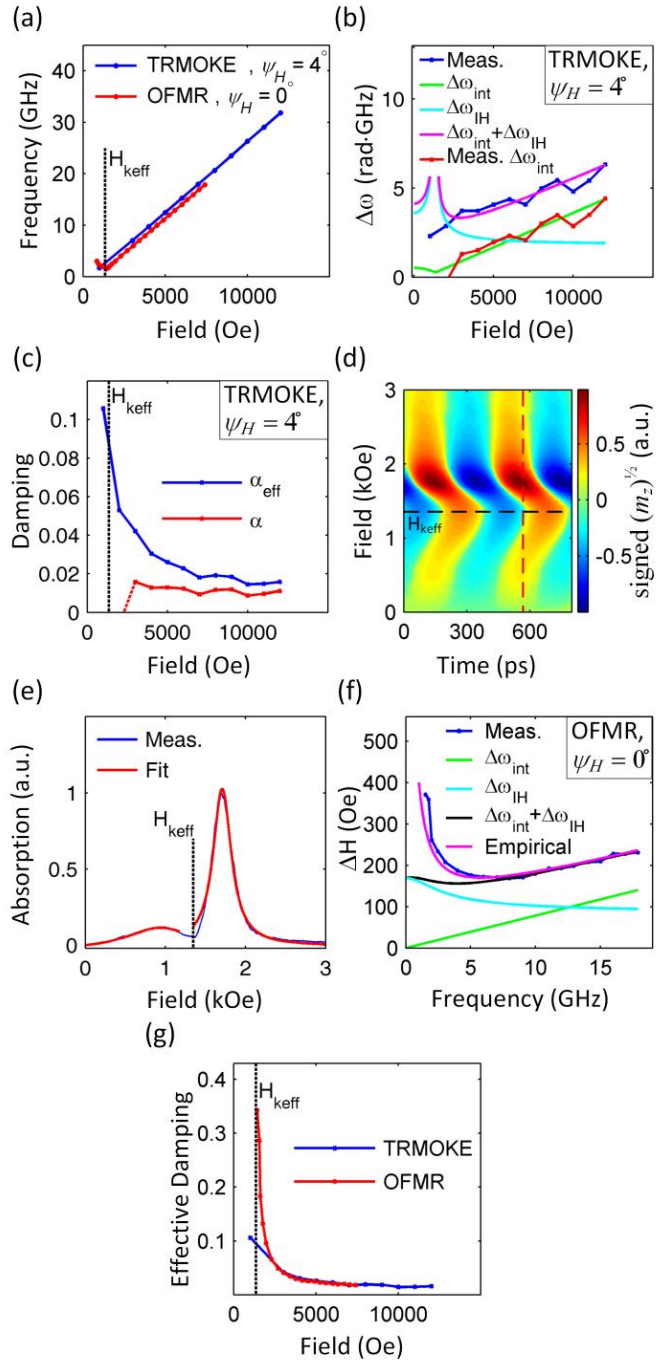

FIG. 4. TRMOKE and OFMR measurements at $\psi_H = 4°$ and $\psi_H = 0°$, respectively. (a) Measured resonance frequency versus magnetic field. (b) Measured linewidth (blue), fitted theoretical contributions to linewidth (green, cyan, magenta) and extracted intrinsic

linewidth from measurement (red) using the TRMOKE with $\psi_H = 4°$. (c) Intrinsic and effective damping using TRMOKE. (d) Representative OFMR trace at 2.5 GHz. The function sign($m_z$)·($m_z$)$^{1/2}$ is plotted. (e) Field dependent absorption spectrum (blue) extracted from the cross section along the red dashed lined of (d) together with fitted lorentzian lineshapes (red). (f) Measured linewidth (blue), fitted theoretical contributions to linewidth (green, cyan, black) and empirical fit that describes the angle misalignment (magenta) using the OFMR with $\psi_H = 0°$. (g) Effective damping using the OFMR and TRMOKE.

Figure 5

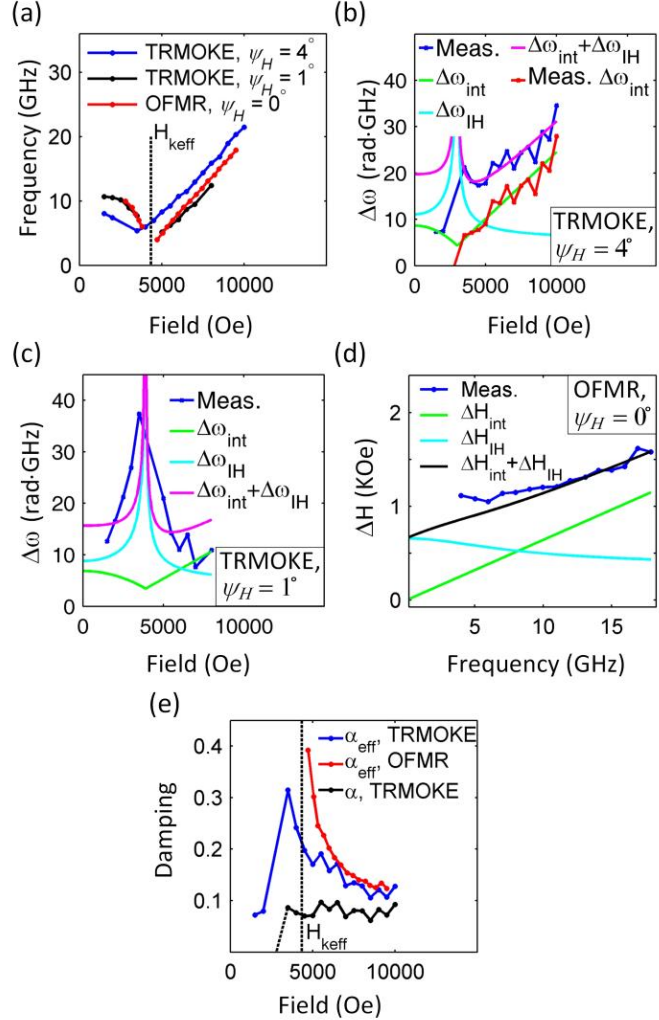

FIG. 5. TRMOKE at $\psi_H = 4°$ and $\psi_H = 1°$ and OFMR measurement at $\psi_H = 0°$ for Co/Ni/Co sample. (a) Measured resonance frequency versus magnetic field. (b) Measured linewidth (blue), fitted theoretical contributions to linewidth (green, cyan, magenta) and extracted intrinsic linewidth from measurement (red) using the TRMOKE with $\psi_H = 4°$. (c) Measured linewidth (blue), fitted theoretical contributions to linewidth (green, cyan, magenta) using the TRMOKE with $\psi_H = 1°$. (d) Measured linewidth (blue), fitted theoretical contributions to linewidth (green, cyan, black) using the OFMR with $\psi_H = 0°$.

(e) Effective (blue) and intrinsic (black) damping using the TRMOKE at $\psi_H = 4°$ and effective damping measured with the OFMR at $\psi_H = 0°$ (red).